# Electronic control of optical tweezers using space-time-wavelength mapping

SHAH RAHMAN,[1] RASUL TORUN,[1] QIANCHENG ZHAO,[1] OZDAL BOYRAZ[1,*]

[1]EECS Department, University of California – Irvine, CA 92697, USA
*Corresponding author: oboyraz@uci.edu

We present a new approach for electronic control of optical tweezers by using space-time-wavelength mapping (STWM), a technique that uses time-domain modulation to control local intensity values, and hence the resulting optical force, in space. The proposed technique enables direct control of magnitude, location, and polarity of force hot-spots created by Lorentz force (gradient force). In this paper, we develop an analytical formulation of the proposed STWM technique for optical tweezing. In the case study presented here, we show that 150 fs optical pulses are dispersed in time and space to achieve a focused elliptical beam that is ~20 μm long and ~2 μm wide. By choosing the appropriate RF waveform and electro-optic modulator, we can generate multiple hot-spots with >200 pN force per pulse.

## 1. INTRODUCTION

Acceleration and trapping of organic and inorganic particles using optical forces, a phenomenon referred to as 'optical tweezing', first demonstrated by Ashkin et al. [1], has found a wide range of applications, especially in the biosciences [2–5]. The key component in optical tweezing was the generation of high optical intensity beams that became feasible with the advent of lasers [6-9]. With coherent laser beams and high numerical aperture lenses, it is possible to induce optical forces on dielectric particles at the focal spot. The optical forces arise due to the transfer of momentum from the scattering of incident photons, and have traditionally been divided into two categories: (a) scattering force, in the direction of light propagation and (b) gradient force, whose magnitude and direction depend on the spatial light intensity gradient [10]. Ashkin's seminal work showed that large enough gradient forces could be achieved using a single laser beam, which could overcome the scattering force, and accelerate particles along an intensity gradient [11]. Apart from single-beam experiments, there has been a growing literature on double beam optical trapping following Ashkin's observations [9, 12-15]. While the scattering force does not lend itself to direct manipulation, there are various ways of manipulating the intensity gradient, and hence of controlling the gradient force, at the focal plane. In the Rayleigh scattering regime in particular, the gradient force is more significant and has been utilized to trap sub-wavelength particles. This opened up possibilities of manipulating the intensity gradient of a laser beam to manipulate particles, leading to applications in microfluidic systems, cell sorting, and characterization of microorganisms. To control the gradient force, the intensity gradient is traditionally manipulated by mechanically steering multiple parallel beams focused onto a narrow region (~50 μm × ~5 μm) [16]. However, to provide greater flexibility and precision, an entirely electronic setup that does not involve any moving parts is more desirable for optical micromanipulation.

In contrast to mechanical beam steering, holographic lenses appeared as an alternative approach to electronically control the intensity gradient at the focal plane and led to the development of holographic optical tweezers (HOTs) [17–21]. In an HOT setup, a computer controlled diffractive optical element (DOE) is used to modulate the phase front of a laser beam before feeding it into a conventional tweezer. Such techniques typically use liquid crystal spatial light modulators (SLM) to create a pixel array of intended phase shifts which are then imposed on the beam at the corresponding pixels. By using different configurations of phase shift arrays, one is able to control the 3-D location of optical traps at the focal volume. Polarization control has also been demonstrated using the phase shift configurations imposed by the SLM, thereby controlling the spin angular momentum of the traps in addition to their spatial location [22]. While HOTs achieve electronic control over a grid of optical traps to some extent, the technique is challenged by the confinement of optical traps to discrete spots as opposed to the ability to manipulate an extended continuous intensity landscape [18]. The lack of direct control over the intensity profile gives rise to inhomogeneity in the intensities of resulting traps, and produces undesired intensity peaks known as 'ghost traps,' which are strong enough to trap particles at unwanted locations [19, 20, 23]. Multiple approaches have been proposed to suppress ghost traps, including random mask encoding [21, 24], reduction of symmetry of the trapping pattern [19], and blocking the zeroth order of diffraction and corresponding ghost traps [20]. Overall, each solution has its pros and cons, and trade-offs exist between various factors such as diffraction efficiency, computational complexity, loss of functionality, power requirement, feasible number of traps, persistent ghost traps etc.

Here, we introduce electronic control of optical tweezing by using space-time-wavelength mapping (STWM) technology [25]. We show that STWM is a powerful approach for creating desired intensity gradients in a 2-D focal plane and for providing direct control of the force field acting on particles. This technique has previously been utilized for real time imaging and detection of micro particles, real time wide field microscopy, arbitrary waveform generation, in photonic time-stretched analog-to-digital converters, waveform digitizers, and in investigation of optical rogue waves [26-35]. In this paper, we build up on our previous study [25], and present further details and applications of the STWM technique in optical tweezing. In this technique, time-domain modulation of power spectral density is used to manipulate the intensity gradient at the focal plane of a laser beam, which in turn controls the 2-D force profile at the focal plane of the beam. Time-domain modulation is then translated onto space domain using a diffraction grating. Using this method, we demonstrate that commercial femtosecond pulses can generate >200 pN optical force hot-spots, whose polarity and location along the focal plane can be controlled to achieve numerous applications such as cell stretching or particle sorting. The amount of force can be controlled by chirped pulse amplification without any nonlinear distortion. Electro-optic modulation of power spectral density allows tuning of local intensity at the focal plane for greater control and to generate attractive and repulsive forces between particles at specific spots as depicted in Fig. 1.

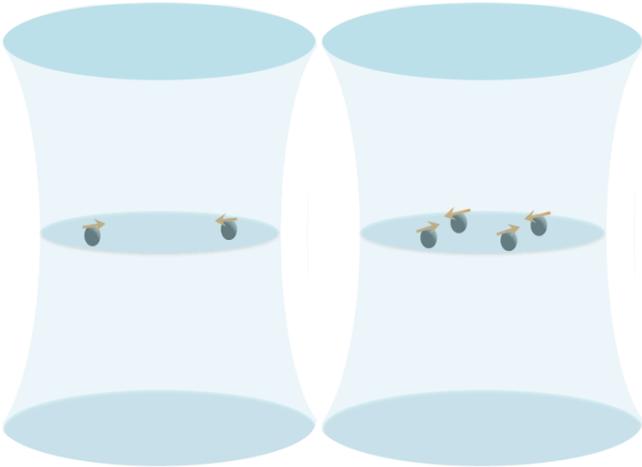

Fig. 1. Particles experiencing gradient force at various spots on the focal plane of the beam (particle size has been scaled up for illustration).

Through the ability to directly manipulate a continuous landscape, this technique has the potential to overcome the limitations of HOTs discussed above, while allowing greater and more direct electronic control of optical traps. STWM eliminates the issue of unwanted inhomogeneity among produced traps, while not compromising efficiency. We also show that the proposed technique can be easily tailored to 1-D and 2-D applications by using diffractive optics such as gratings and virtually-imaged phased arrays (VIPA) [36].

The paper is organized as follows: Section 2 describes the experimental model used in simulations and presents a theoretical analysis based on each component in the model. Section 3 presents simulation results and describes techniques to control the location, polarity, and size of the optical force to suit various applications. And Section 4 gives a summary of the methods presented and discusses some limitations and challenges for future work.

## 2. MODELING SETUP AND THEORY

To demonstrate the feasibility of optical tweezing by using space-time-wavelength mapping (STWM), we develop our analysis based on the following realistic experimental setup as illustrated in Fig 2. The optical tweezer is driven by a fiber laser at 1550 nm that produces 150 fs pulses with ~20 kW peak power. These values correspond to a laser with ~100 mW average power at 50 MHz repetition rate. The laser beam then propagates through a dispersive medium with ~100 ps/nm chromatic dispersion to broaden the optical pulse width to ~2 ns at the full width half max (FWHM) point. Hence, the dispersive propagation creates a time-wavelength mapping where 1 nm wavelength separation corresponds to 100 ps separation in time domain. Dispersed pulses are then passed through an electro-optic modulator to manipulate the power spectral density of the laser as desired, by using RF waveforms. Here we assume no losses in components and that nonlinear distortion is absent in dispersive fiber. In practice, it is possible to start with a lower peak power and add a chirped pulse amplification to compensate losses and eliminate distortion. The time-wavelength mapped pulses are then coupled to a diffraction grating with 600 lines/mm groove density and a cylindrical lens to create space-wavelength mapping. Finally, we use a 60x microscope objective to generate a ~20 μm long and ~2 μm wide elliptical focal spot. After time-wavelength mapping, the diffraction grating provides a space-wavelength mapping of 1 μm/nm at the focal plane.

Through STWM, we obtain a desired wavelength-dependent intensity profile of $I(\lambda(x,t))$. We develop optical force calculations based on this intensity distribution at the focal plane ($z = 0$). In particular, we only assume optical forces in the lateral direction and ignore the force components in the axial direction. We start with a Gaussian beam $A(t) = A_0 \exp\left[-\frac{1}{2}\left(\frac{t}{T_0}\right)^2\right]$. Through time-wavelength mapping and RF modulation, we generate a field profile of $\boldsymbol{E}(t) = A_{\text{disp}}(t) e^{j\omega t} e^{j\phi(t)} f_{\text{RF}}(t)$ where $A_{\text{disp}}(t)$ is the dispersed pulse amplitude, $\omega$ is the angular frequency, $\phi(t)$ is the time-dependent phase or chirp, and $f_{\text{RF}}(t)$ is the applied RF modulation. This field corresponds to the optical power profile,

$$\boldsymbol{P}(t) = \boldsymbol{P_{in}}(t) f_{RF}^2(t) = \boldsymbol{P_{in}}(t)\left[\frac{1}{2} + \frac{1}{2}\cos\left(\frac{\pi}{V_\pi}\left(V_{bias} + v_{RF}(t)\right)\right)\right], \quad (1)$$

where the input power $P_{\text{in}}(t) = (A_{\text{disp}}(t))^2$, $V_\pi$ is the half-wave voltage of the modulator, $V_{\text{bias}}$ is the DC bias voltage, and $v_{\text{RF}}(t)$ is the applied RF waveform. By manipulating $v_{\text{RF}}(t)$ using an arbitrary waveform generator, we can control the power spectral density of the laser emerging from the electro-optic modulator.

The aggregate intensity profile of all wavelengths after the space-wavelength mapping, i.e., after the diffraction grating and a Fourier lens, can be formulated as [29].

$$I(x,\lambda) = \sum_i \frac{P_i}{(f\lambda_i)^2} exp\left[-\left(\frac{2\pi w(x-x_{pi})}{f\lambda_i}\right)^2\right] \quad (2)$$

Here, $w$ is the spot size of the beam, $f$ is the focal length of the Fourier lens, $P_i$ represents the peak power carried by wavelength $\lambda_i$ in the spectrum, $x_{pi} \approx G_\beta f(\lambda_i - \lambda_c)$ is the relative position of the first order diffraction peak for the wavelength $\lambda_i$ with respect to the central wavelength $\lambda_c$. $G_\beta = \frac{G}{\cos(\beta)}$ is the effective groove density of the diffraction grating shown in Fig. 2, that is defined as a function of groove density $G$ and the first order diffraction angle $\beta$. For a beam with wavelength $\lambda$, and incident angle $\alpha$, the first order diffraction angle $\beta$ is given by the grating equation $\sin(\alpha) + \sin(\beta) = \lambda G$ which reduces to $2\sin(\beta) = \lambda G$ when $\alpha = \beta$ in a Littrow configuration [29].

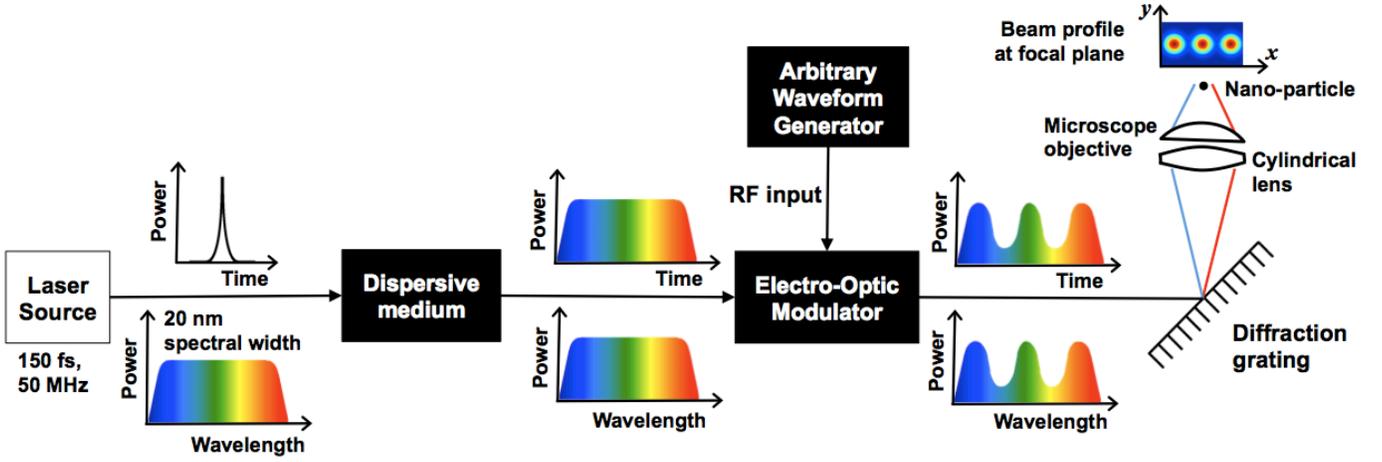

Fig. 2. Proposed experimental setup. A 150 fs laser with 50 MHz repetition rate undergoes dispersion in time and is then modulated in time domain, and simultaneously in wavelength domain, using arbitrary RF waveforms fed to an electro-optic modulator. The modulated power spectral density is then mapped onto space domain using a diffraction grating. A cylindrical lens and a microscope objective then focus the diffracted beam onto a tight focal spot (20 x 10 μm) for optical manipulation of particles.

To develop a formulation for the optical force at the focal plane, we assume operation under the Rayleigh scattering limit and that the particles are polarizable, such as dielectric particles. The gradient force from a laser beam acting on a particle is proportional to the light intensity gradient and it can be derived from the basic formulation of Lorentz force [37].

$$F = \frac{1}{2}\alpha \nabla E^2 \quad (3)$$

Incorporating proportionality constants, we get,

$$F(r) = \frac{2\pi n_0 b^3}{c}\left(\frac{m^2-1}{m^2+2}\right)\nabla I(r) \quad (4)$$

where $m = n_1/n_0$ is the relative refractive index of the particle (where $n_1$ is the refractive index of the particle and $n_0$ is the refractive index of the surrounding medium), $b$ is the radius of the particle, $c$ is the speed of light, and $r$ is the radial distance from the beam axis.

Taking the gradient of intensity in (2) with respect to the radial distance, we get (5):

$$\nabla I(x) = \frac{\partial I(x)}{\partial x}$$
$$= -\sum_i \frac{8\pi^2 w^2 (x-x_{pi})}{(f\lambda_i)^4} P_i \, exp\left[-\left(\frac{2\pi w(x-x_{pi})}{f\lambda_i}\right)^2\right] \quad (5)$$

Here we have assumed that we are at the focal plane and that the beam width is the spot size at the focus, *i.e.*, $w = w_0$ for $z = 0$, $z$ being the direction of propagation of the laser. Substituting (5) in (4), we get an expression for the aggregate optical force along $x$, as shown in (6):

$$F(x) = -\frac{16 n_0 w^2 \pi^3 b^3}{c}\left(\frac{m^2-1}{m^2+2}\right)$$
$$\times \sum_i \frac{(x-x_{pi})}{(f\lambda_i)^4} P_i \exp\left[-\left(\frac{2\pi w(x-x_{pi})}{f\lambda_i}\right)^2\right] \quad (6)$$

Note that in equation (6), we have defined the positive $x$ direction to be the direction of positive force. Also, equation (6) represents a direct relationship between the applied RF modulation and the force profile at the focal plane. RF modulation controls the power spectral density $P_i$ in (6), thus modulating the beam intensity profile, and allowing direct control of the force profile $F(x)$.

## 3. RESULTS AND APPLICATIONS

Fig. 3 illustrates the intensity profiles and optical force contours acting on particles for three different RF modulation signals. The modulator is biased at the quadrature point and supplied with RF pulses of peak-to-peak voltage $v_{RF\_PP} = \gamma \cdot V_\pi$, where the amplitude factor $\gamma$ is controlled by the arbitrary waveform generator. In the first row of Fig. 3, we present the case when there is no RF signal (Fig. 3a). As expected, we generate a pure elliptical beam that produces a force contour that traps the particle along the central axis of the beam. Effectively, for a beam of size 20 x 2 μm, the particle is trapped along a line-shaped well with dimensions 15 x 1.5 μm as shown in Fig. 3b and 3c. In the second row, we apply a 2 ns wide pulsed RF signal where peak-to-peak voltage $v_{RF\_PP} = 1V_\pi$ (Fig. 3d). In the second row, the RF pulse wave in Fig. 3d creates two 3 x 2 μm intensity spikes separated by a 4 μm null region as depicted in Fig. 3e. Each intensity spike creates a 4 x 1 μm trapping region as illustrated in Fig 3f. Similarly, Figures 3g, 3h and 3i in the third row illustrate the RF waveform and generation of three intensity spikes and three force hot-spots. Here, the RF waveform has two pulses, each of width 1 ns, separated by 2 ns (Fig. 3g). It cuts through the beam at two places, leaving three intensity spikes, each measuring 2 x 2 μm (Fig. 3h). These three intensity spikes in turn generate three force hot-spots, each of which can trap particles within a 4 x 1 μm region. Beams with wider spectra can be used to produce higher number of hot-spots along the ellipse of the focus. The maximum force generated per pulse at the hot-spots is ~200 pN for 100 nm particle radius. By using different waveforms we can extend the number of hot-spots, and control the relative strength and position of optical forces acting on polarizable particles. In these calculations, we assume that the rise time of RF pulses is >200 ps, which requires the RF bandwidth of the arbitrary waveform generator to be well below 10 GHz.

One of the clear advantages of the proposed method is the precise location control of force fields without altering the rest of the continuous force landscape. The center of the force field is determined by the location of the intensity spikes. Hence, we can change the pulse width and RF voltage to determine the exact location of trap centers. For instance, for the RF signal shown is Fig. 3d, a pulse width $\tau_1 = 2$ ns corresponds to a distance $d_1 = 6$ μm between force hot-spots, and a pulse width $\tau_2 = 4$ ns corresponds to a distance $d_2 = 12$ μm as illustrated in Fig. 4, by the red and blue curves respectively. The precision of this technique depends on the RF bandwidth. For a bandwidth of 10 GHz, the narrowest RF pulse possible is <0.1 ns, which results in a minimum achievable separation distance of <0.5 μm. This means that two particles separated by, for example, 5 μm, can be brought to within 0.5 μm of each other using this approach.

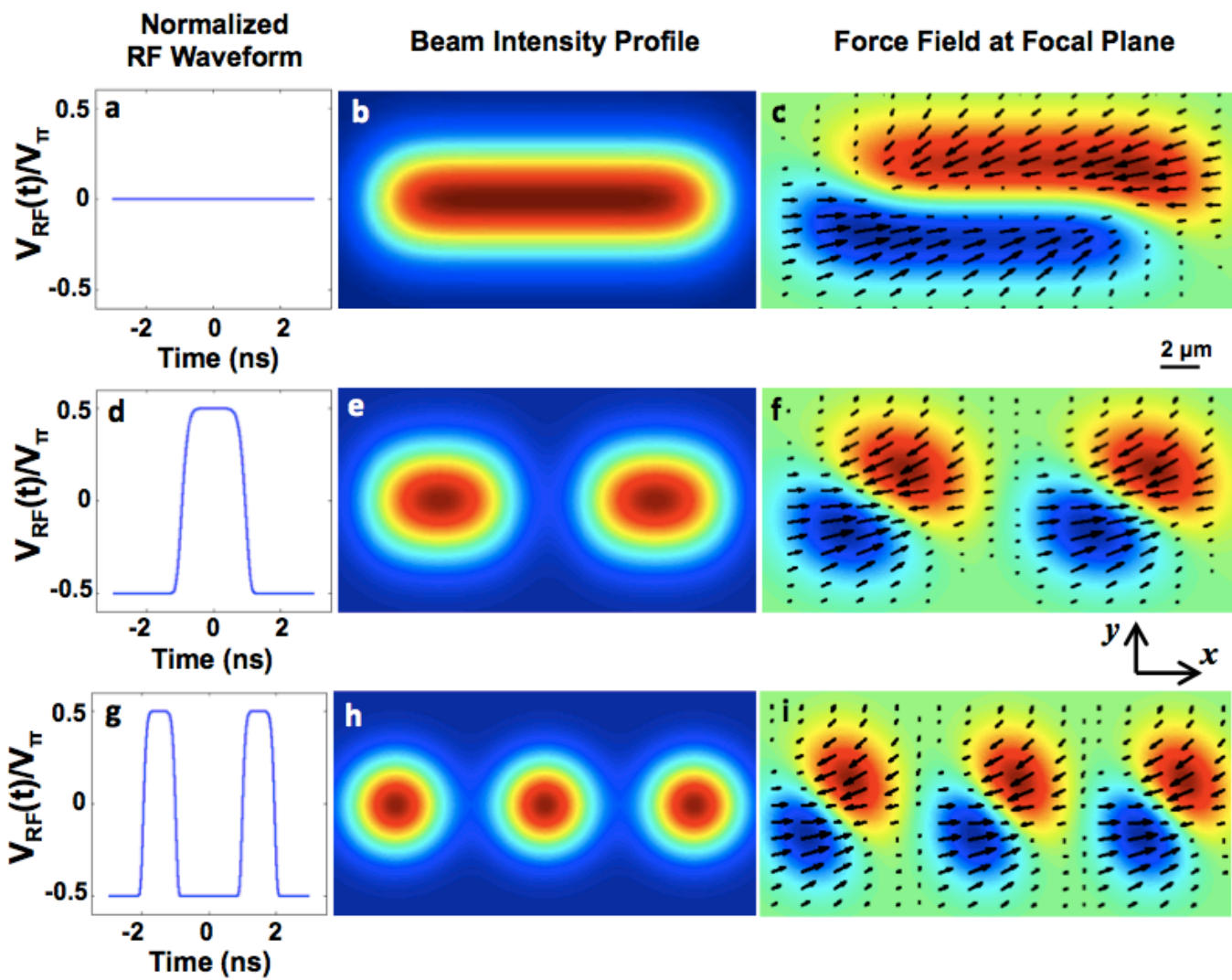

Fig. 3. Different RF waveforms create different intensity profiles, and hence can be used to produce different 2-D force profiles. Figures on the left column show RF waveforms used to modulate the intensity. The middle column shows intensity profiles corresponding to each RF waveform, and the right column shows corresponding force contours arising from each intensity profile. Hot-spots or points of maximum force can be maneuvered over the focal plane by selecting an appropriate RF waveform.

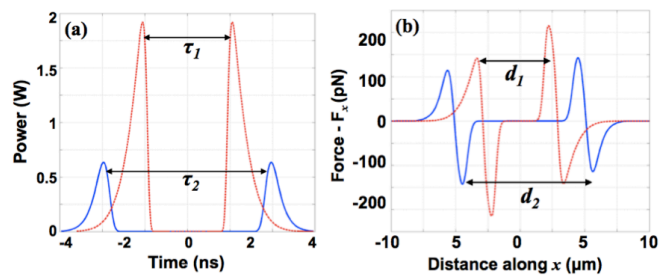

Fig. 4. Illustrating location control of hot-spots. (a) Modulated power distribution, (b) Generated force profile along $x$.

In addition to location control of hot-spots, Fig. 5. illustrates how we can also alter the direction of the force at the hot-spots by generating suitable intensity profiles through RF modulation. To achieve this, we need the ability to change the sign of the intensity gradient. A default super-Gaussian beam profile (Fig. 5a) results in a trap located at the beam center (Fig. 5c) due to a positive intensity gradient, followed by a negative intensity gradient (Fig. 5b). As depicted in Fig 5b, a positive gradient (~ +1.5 W/nm) produces a positive force spike (~250 pN), followed by a zero-force region spanning ~5 μm. Then a negative gradient (~ -1.5 W/nm) creates a negative force spike (~ -250 pN). A particle positioned at each of these spikes will experience a force pulling them toward the center, creating an attractive force between the particles. Such pulse profiles are particularly useful for bio applications that compress cells. In general, the precise control of DC bias and RF waveform facilitates the generation of arbitrary force polarity and magnitude.

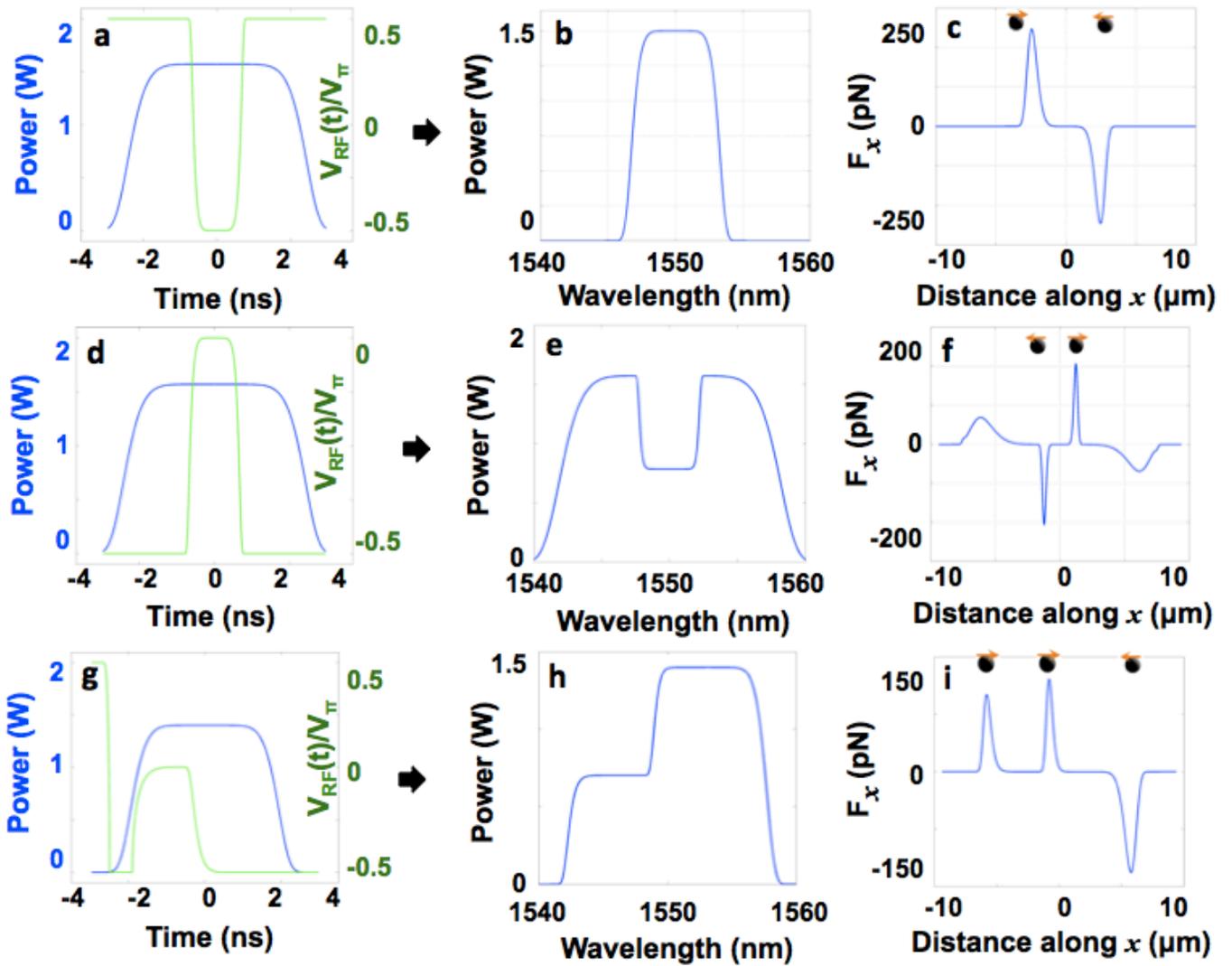

Fig. 5. Demonstrating polarity control of hot-spots. The green curves on the left column show the normalized RF waveform; the middle column shows the corresponding modulated power profile for each RF waveform, and the right column shows the desired force profiles generated using the applied modulation. Different sequences of positive and negative power gradients result in different arrangements of polarities (i.e. push or pull) of the generated force spikes, creating attractive (top row), repulsive (middle row), and other arbitrary (bottom row) force configurations.

To reverse the force direction, we can modulate the intensity at the beam center and create negative gradients. For instance, a pulsed RF waveform such as the one shown in Fig. 5d., can be used to create a power dip in the middle (Fig. 5e), essentially producing a negative gradient followed by a positive gradient at the beam center. This results in an inverted force profile (Fig. 5f compared to 5c), where a negative force spike is followed by a positive one, with a zero-force region in between. In this particular example, we used a 2 ns wide pulsed RF waveform with a peak-to-peak amplitude of $0.5V_\pi$ to create a negative gradient of ~-0.5 W/nm followed by a zero gradient and then a positive gradient of ~+0.5 W/nm. Fig. 5f shows the resulting force profile, which is composed of a negative force spike (~-190 pN) and a positive force spike (~+190 pN), separated by a zero-force region spanning ~4 μm. The configuration in Fig. 5f can be used to repel particles away from the center of the beam. Similarly, we can use more complicated RF waveforms to generate compressive and repelling forces or control multiple particles at the same time as shown in Fig. 5g, 5h and 5i. Here, we use the RF waveform shown in Fig. 5g to modulate the power spectral density such that the power rises in two discreet steps, creating two successive positive gradients. First, the peak power is increased from 0 W at ~1541 nm to a steady value of ~0.7 W at ~1543 nm, which is maintained until ~1549 nm as depicted in Fig 5h. The power profile in Fig. 5h starts with a positive gradient of ~+0.35 W/nm, which produces a positive force peak of ~+140 pN, followed by a zero-force region spanning ~4 μm as shown in Fig. 5i. Then, instead of a power fall off as characteristic of regular pulses, the peak power undergoes a second increase from ~0.7 W at ~1549 nm to ~1.4 W at ~1550 nm, with a gradient of ~+0.5 W/nm, followed by a steady region where it remains at 1.4 W until ~1556 nm, before sliding down to zero with a negative gradient of ~-0.47 W/nm as illustrated in Fig. 5h. Thus, we obtain an additional positive force peak of ~+150 pN due to the second rise in peak power, followed by a zero-force region of width ~6 μm and a negative spike of ~-150 pN as depicted in Fig. 5i.

The proposed technique can be extended to create a 2-D grid of traps or any other arbitrary 2-D pattern by treating the power spectral distribution as a 2-D matrix and modulating each row of the matrix with a desired RF waveform. Another way to create and control a 2-D force field is to use a 2-D spatial disperser to generate a 2-D spectral shower using two orthogonally oriented 1-D spatial dispersers: a diffraction grating and a virtually-imaged phased array (VIPA), as prescribed in literature [36]. The angular dispersion by a VIPA is 10–20 times larger than those of common diffraction gratings, which have a Blaze angle of ~30º [38]. The VIPA employs a thin plate of glass and a semi-cylindrical lens (C lens); the input light is line focused with the semi-cylindrical lens into the glass plate. The collimated light then emerges on the other side of the plate, where the angle of propagation is dependent on the wavelength [38]. When it comes to 2-D manipulation, holographic optical tweezers are typically limited to a grid of optical traps. While our approach has no such limitation, it is useful to mimic such a grid for existing applications. Using the proposed technique, each trapping region and its associated force spikes take up a minimum area of 7 x 4 µm, meaning a ~1200 µm² area can accommodate ~45 traps as shown in Fig. 6. In addition to a rigid grid of traps (Fig. 6), we can achieve any arbitrary pattern with extended areas of high intensity, leading to bigger force hot-spots, whose location and polarity can be precisely controlled over the entire 2-D plane as described above and illustrated in Fig. 7. This opens up new possibilities and further prospects stemming from the greater flexibility and control over the entire landscape.

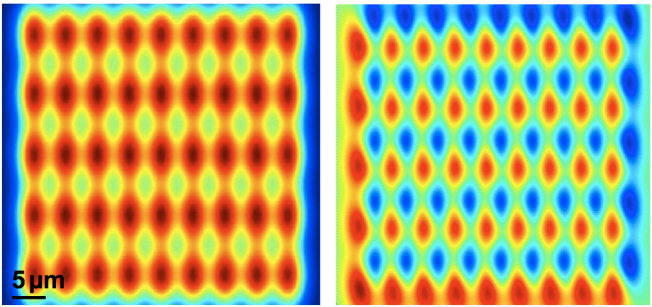

Fig. 6. Using a 2-D extension of the proposed technique, one can generate the beam profile on the left, which can be used to obtain the 2-D grid of optical traps on the right.

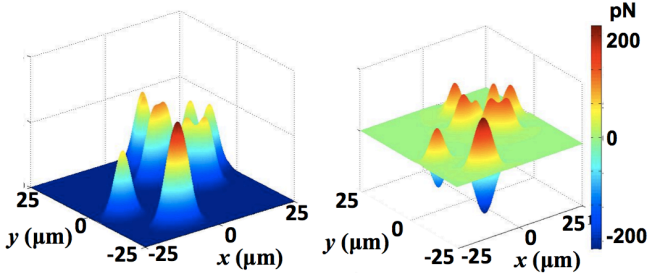

Fig. 7. An arbitrary 2-D force profile (right) may be generated using a corresponding intensity profile (left).

## 4. DISCUSSION AND SUMMARY

We have presented a method to achieve electronically controlled force profiles for optical tweezing applications by extending the idea of STWM. The proposed approach can generate a continuous force profile at the focal point where polarity, location and magnitude can be controlled precisely by using time-domain electro-optic modulation. The created hot-spots can be actively tuned to accelerate particles in a particular direction, to maneuver them along the 2-D focal plane, and to create more complex scenarios for a wide range of applications, especially in the biosciences.

One major obstacle that continues to be a challenge in optical manipulation is the diffraction limit when working with nanometer-sized objects. Techniques in plasmon nano-optics exploit surface plasmon resonances supported by metallic nanostructures, to effectively concentrate evanescent fields well beyond the diffraction limit [39-48]. The STWM technique proposed here, however, uses time-stretch to overcome the diffraction limit. Precision of the proposed technique can be enhanced by using a wider RF bandwidth, or by increasing the optical bandwidth of the laser with an appropriate dispersion value. Overall, one needs to consider trade-offs between dispersion, RF bandwidth, spectral width, power requirement, and spatial precision, in order to implement the STWM approach. Future work building up on this study will involve conducting the proposed experiment to corroborate the theory and simulation with experimental results, and further investigation into the capabilities and limitations of the proposed method. In particular, the analysis will be extended by considering scattering forces in the axial direction, which have been ignored in the current model.